\documentclass{aa}  
\usepackage{graphicx}
\usepackage{amsmath}    
\usepackage{amssymb}    
\usepackage{times}
\usepackage{url}
\usepackage{color,graphicx}
\usepackage{txfonts}
\definecolor{czer}{rgb}{0.99,0.0,0.0}
\definecolor{ziel}{rgb}{0.0,0.8,0.2}
\newcommand{\zi}[1]{\colorbox{ziel}{#1}}
\newcommand{\cz}[1]{\colorbox{czer}{#1}}
\begin{document} 

   \title{Remarks on generating realistic synthetic  meteoroid orbits}

   \author{T. J. Jopek
          \fnmsep\thanks{jopek@amu.edu.pl}
          }

   \institute{Astronomical Observatory Institute, Faculty of Physics, A. Mickiewicz University, ul. Sloneczna 36, 60-286 Pozna\'{n}, Poland
             }

   \date{Received ???; accepted ???}
   
  \abstract
   {To identify the real associations of small bodies, we can use synthetic sets of orbits generated by various methods. These are not perfect methods, therefore the assessment of their quality is an essential task.}
   {In this study, we compared five methods for generating synthetic meteoroid orbits. Three of them (ME0, KD10, and KDns) had already been proposed in the literature, while two additional ones (ME1 and ME4) are new methods.}
   {As far as possible, the synthetic orbits were compared with the orbits of the observed meteoroids. For quantitative comparison, we applied a few tests: the $\chi^2$-distance and{ the nearest neighbor} $NN_N$ tests used in { previous works}, and one-dimensional $\chi^2$ and Kolmogorov-Smirnov (K-S) tests, as well as a two-dimensional  K-S test implemented in this study. To estimate a general property of the orbital sample, we proposed the use of the  entropy $H_N$ of the data set based on the nearest neighbor distances. Finally, we did a cluster analysis of the synthetic orbits. We calculated and compared the values of the orbital similarity thresholds.}
   {We showed that generating 'realistic' meteoroid orbits and testing their quality is a complex issue. An assessment of the quality of the generated orbits depends on the type of test applied, and it refers to the sample of the observed orbits used. Different tests give different assessments. However, in practice, the investigated methods produced similar results if they were applied correspondingly.}
   {}
   \maketitle
%

\section{Introduction}
\label{intro}
Searching for streams among meteoroid orbits, including an assessment of their reliability,  is based heavily on the artificial orbital samples derived by means of the statistical properties of the observed orbital elements. 
One can derive the synthetic orbital samples, free from clusters, by different methods (see, for example,  \citet{1997A&A...320..631J}, \citet{2003MNRAS.344..665J}, \citet{2005M&PS...40.1241P}, \citet{2014Icar..239..244K}, \citet{2019A&A...622A..84G}). 
Recently, \citet{2017P&SS..143...43J} have shown that the choice of the method used for generating synthetic orbits impacts the assessment of the probability of spurious pairings. 
\citet{2017P&SS..143...43J} tested a few methods and, in the case of meteoroid orbits, they recommend { their} method as the most reliable. However, \citet{2017Icar..296..197V} found some drawbacks of this method, and concurrently they proposed an interesting approach based on the Kernel Density Estimation method { (KD method)}. 

In this study, we describe an improvement of method E, by eliminating the drawbacks pointed out by \citet{2017Icar..296..197V}, and by introducing into the algorithm a few additional steps that more precisely allow for the correlations between the meteoroid orbital elements. The modified methods have been compared with the KD methods proposed by \citet{2017Icar..296..197V}.
%

\section{Orbital data used in this study}
%
\label{dataused}
In this study, the same meteoroid data  have been used as in \citet{2017Icar..296..197V}, namely, the file CAMS-v2-2013.csv
\citep[see ][]{2016Icar..266..331J} downloaded from the GitHub site (\url{https://github.com/dvida/GenMeteorSporadics}). 
We used the heliocentric ecliptical osculating orbital elements $q, e, \omega, \Omega,  i$ and the ecliptic latitude  $\beta_G$ of the meteor geocentric radiants. Additionally, we use the heliocentric distances of the Earth $R_E$, calculated  at the moment of meteor observations by means of the formula taken from \citet{2012asal.book.....B}.
Using the software posted by \citet{2017Icar..296..197V} at the GitHub site, from the original CAMS file, we extracted $58029$ orbits of sporadic meteors, which we considered to be an orbital sample free from clusters.  
%
\section{Synthetic meteoroid orbits generation methods}
In this study, we compared a few methods: ME0  (method E in \citet{2017P&SS..143...43J}) and its two variations ME1 and ME4. The ME methods use the CPD (cumulative probability distribution) inversion technique parametrized by the choice of the bin width of the histogram used for each orbital element. In the ME0 method, it is assumed that the Earth moves in a circular orbit around the Sun.%
{ The ME1 method is based on the ME0 method, in which the assumption of the elliptical orbit of Earth was introduced. In turn, the ME4 method is an extended version of the ME1 method. The values of the argument of perihelion  $\omega$ were generated by the CPD technique using not one marginal distribution but two, corresponding to the positive and negative ecliptic latitudes of the meteor radiants.}

Additionally, we made use of three KD methods:  KD0.1, KD10, and KDns (see, \citet{2017Icar..296..197V}).
The KD methods use the nonparametric technique; they are based on the multivariate kernel density estimation. The methods depend on the choice of the bandwidth matrix $H$. This matrix has a decisive influence on the shape (smoothness) of the obtained multivariate probability distribution function. 
In the work of \citet{2017Icar..296..197V}, three diagonal H matrices were used, two scalar matrices with elements equal to $0.1$ and $10$ (methods KD0.1 and KD10 respectively) and one non-scalar matrix (KDns method) with elements selected separately for each orbital element. More details about the applied ME and KD methods are provided in the Appendix.
\section{Valuation of the synthetic orbital sample}
\subsection{Comparison of the observed and synthetic samples}
\label{sec:comparisonmethods}
The ME method uses the marginal densities represented by the marginal histograms of the observed orbital elements{ (see Fig.~\ref{fig:HiCAB})}. The application of the CPD inversion technique guarantees that the marginal distributions of the synthetic and observed orbital elements are very similar. However, this technique does not conserve the correlations between the meteoroid orbital elements $q,e$, and $\omega$  (see Fig.~\ref{fig:2DCAB}). To address these correlations, \citet{2017P&SS..143...43J} added steps 2 and 3 to the ME0 method (see Appendix \ref{ME}).  However, in their paper no comparisons between the obtained synthetic and observed samples were accomplished, except for a visual assessment.

In  \citet{2017Icar..296..197V}, to ascertain the differences or similarities between the observed and synthetic orbital sets, a few comparisons were accomplished, namely,  
a visual comparison of 1D or 2D histograms of the observed and synthetic orbital elements, and a limited quantitative assessment of the similarity between observed and synthetic samples. For this purpose, \citet{2017Icar..296..197V} used the 2D histogram of $\omega$ versus $q$, for which the so-called $\chi^2$ histogram distance was calculated \citep[see][]{2010confP} using
 \begin{equation}
 \chi^2(P,Q)=0.5\sum_{i=1}^{k_{\omega}} \sum_{j=1}^{k_{q}} \frac{(P_{ij}-Q_{ij})^2}{(P_{ij}+Q_{ij})},
 \label{Pele}
\end{equation}
 where $P_{ij}$ and $Q_{ij}$ are the contents of the $ij$ bins of the synthetic and observed data histograms, respectively. The number of bins are represented by $k_\omega$ and $k_q$  .
 
 In \citet{2017Icar..296..197V}, as an additional statistical property of the sample, the arithmetic mean  ${NN_N}$ of the nearest neighbor distances  was calculated, which is the averaged minimum distance from each orbit to any other for a given sample. Let $D_{N,i}$ be the nearest neighbor distance of the orbit $O_i$ and the other $O_j$: $D_{N,i}=\min_{j\neq i, j\leq N}\,D(O_i,O_j)$, then
 \begin{equation}
 {NN_N} = \frac{1}{N}\sum_{i=1}^N D_{N,i}
 \label{DNN}
 ,\end{equation}
 where $N$ -- is the sample size. To determine the distance  $D(O_i,O_j)$  between two orbits, the $D_{SH}$-function was applied \citep[see,][]{1963SCoA....7..261S}.
 
 As a statistical property between two orbital samples, \citet{2017Icar..296..197V} calculated the arithmetic mean ${OS_N}$ of the smallest D-values between each synthetic orbit paired with each observed orbit.
 
 To compare the two orbital samples, in the present work  we have used ideas from \citet{2017Icar..296..197V}, expanding them slightly, as well as additional new ideas,  namely  
 we compared the marginal distributions of $q,e,\omega,\Omega, i$ as well as the 2D distributions of all pairs of orbital elements using the $\chi^2$ and  Kolmogorov-Smirnov (K-S) tests taken from \citet{2002numrec.book.....P}. The advantage of this approach is the ability to estimate whether the compared samples could come from the same population.
 
 To assess the general similarity of orbital samples, we used the concept of entropy. As far as we know, this concept has never been used for this purpose. The entropy $H_{N}$ was calculated on the basis of the nearest neighborhood values $D_{N,i}$. For each point in the sample, one has to find the distance to its nearest neighbor and apply the formula \citep{1997IJMSS....6...17B}
 \begin{equation}
H_{N}=\frac{1}{N}\sum_{i=1}^N  \left [ \ln(N\cdot D_{N,i}) \right ] +\ln(2) + C_E
\label{HN}
 \end{equation}
where $C_E=0.577 215 664 901 532 860 ... $ is the Euler constant.  
An analogy between ${NN_N}$ and $H_N$ is clearly seen.

 Finally, we propose a purely practical approach. For the synthetic orbits, we compared the orbital similarity thresholds used in the cluster analysis. We determined the thresholds by a single linkage cluster analysis and three $D-$functions. More details of this procedure are given in \citet{2020MNRAS.494..680J}. 
 
In the next section, we present the results of our calculation of the $\chi^2$ distances and ${OS_N}$, ${NN_N}$ statistics, as well as the results of the $\chi^2$ test and K-S test, using the observed and synthetic data sets. Moreover, we present the values of the entropy and orbital similarity thresholds calculated for the synthetic orbital samples.
\subsection{Results of comparison and discussion} 
In this section, we present the results of quality tests on synthetic sets of $58029$ and $29014$ orbits. First, tests carried out by \citet{2017Icar..296..197V} were repeated. The following sections present the results of the extended comparison of reduced orbital samples. The last section describes the results of the cluster analysis among the generated orbits.
\subsubsection{Samples comprising 58029 orbits}
\label{fulldatacomparison}
At this stage, we only recalculated the results already obtained by \citet{2017Icar..296..197V} that are provided in their Table~1. 
Exploiting the software placed by Vida et al.\footnote{Actually we used the corrected version of this software. A few corrections and  a small supplementation were implemented by Denis Vida.} on the GitHub website, 
we generated four synthetic samples and calculated their  $\chi^2$ distances and ${OS_N}$,  ${NN_N}$ statistics (see our Table~\ref{tab:comparison1}).
\begin{table}
\scriptsize
\footnotesize
\caption[]{
The $\chi^2$ distances and ${SO_{N}}$, ${NN_{N}}$ statistics calculated for $58029$ orbits, observed and generated by the ME method and three variations of the KD method, see text.
In the second column, the $\chi^2$ distances  were calculated using $\omega$-$q$ histograms with $158$x$103$ bins. Additionally, in the third column the $\chi^2$  distances were determined using $80$x$80$ { bins}, see text.
}
\begin{center}
\begin{tabular}[t]{lcccc}
\hline
\hline
\multicolumn{1}{l}{Orbital sample} & \multicolumn{2}{c}{$\chi^2$ distance} & \multicolumn{1}{c}{${OS}_{N}$} &  \multicolumn{1}{c}{${NN}_{N}$}\\
\multicolumn{1}{l}{} & \multicolumn{1}{c}{158x103}& \multicolumn{1}{c}{80x80} &  & \\ 
\hline
CAMS  58029      &         -    &     -    &      -   &  0.0806 \\ 
ME               &     19053    &   16285  &  0.0995  &  0.1008 \\
KD10           &     16376    &   14048  &  0.0821  &  0.0896 \\
KD0.1          &     1858     &    933   &  0.0408  &  0.0447 \\
KDns           &     9189     &    6921  &  0.0776  &  0.0840\\
\hline
\label{tab:comparison1}
\end {tabular}
\end {center}
\normalsize
\end {table}

The $\chi^2$ distances given in the second column of Table \ref{tab:comparison1} were obtained using $q$ versus $\omega$ histograms with $158$x$103$ bins,{ as used in \citet{2017Icar..296..197V}}.
Our results differ slightly from those found by \citet{2017Icar..296..197V} due to two reasons: we used corrected software, and we used a different sequence of random values starting with the seed point $2003$.

As in \citet{2017Icar..296..197V}, in Table~\ref{tab:comparison1} the largest $\chi^2$ value was obtained for the ME sample, the smallest value for the KD0.1 sample.
However, we must point out that the $\chi^2$ distances given in Table~\ref{tab:comparison1} have no clear interpretation. The high value $\chi^2$=$19053$ does not necessarily mean that in some applications the ME synthetic sample cannot effectively represent the observed orbits. 
On the other hand, one may ask, does the smallest value $\chi^2$=$1858$ indicate that the synthetic sample generated by the KD0.1 method is too similar to the observed one?

Furthermore, the $\chi^2$ distance considerably depends on the number of applied bins. This is shown in the third column of Table~\ref{tab:comparison1} where  we placed the $\chi^2$ distances calculated using histograms with $80$x$80$ bins. The number of bins was calculated according to the Rice Rule: $k=\lceil 2 N^{1/3} \rceil$, where $\lceil \ldots \rceil$ is the ceiling function and $N$ is the sample size. 
The obtained $\chi^2$ distances differ significantly from those given in the second column of Table~\ref{tab:comparison1}. Hence, the question may arise, which $\chi^2$ value should be used to valuate the quality of a synthetic sample?
Thus, we believe that the $\chi^2$ value determined by Eq. (\ref{Pele}) should be used with care for the adjudication of a  synthetic orbits generator. 

In \citet{2017Icar..296..197V}, the authors used another statistics for a quantitative comparison of the observed and synthetic samples. As was mentioned in Sect. \ref{sec:comparisonmethods}, the authors proposed two statistics: the arithmetic means ${NN_N}$ and ${OS_N}$.
We also calculated ${NN_N}$ and ${OS_N}$ and the obtained results are given in Table~\ref{tab:comparison1}. Our results differ slightly from those given in Table~1 in \citet{2017Icar..296..197V} for the same reasons as those we mentioned earlier. 
The highest values of ${NN_N}$ and ${OS_N}$ are for the ME sample. \citet{2017Icar..296..197V} considered the ${OS_N}=0.0995$  as an upper bound for an acceptable synthetic orbit data set. As the lower bound, Vida et al. considered a value of  ${OS_N}=0.0408$. 
Certainly, a low value of ${OS_N}$ may imply a strong similarity between the two samples. However, we are interested in a statistical similarity. Therefore, as in the case of the $\chi^2$ distance, one may ask, which limit values of ${OS_N}$, ${NN_N}$ may be relevant to the issue of the comparison between synthetic and observed samples?
We address this question in the next section.
\subsubsection{Samples comprising 29014 orbits }
\label{camsab}
To find the expected values of the $\chi^2$ distances, ${OS_N}$ and ${NN_N}$ , and other statistics,  we split the original CAMS sample into two subsamples, CAMSA and CAMSB, each containing $29014$ orbits. The CAMS sample is sorted { by increasing values of the ecliptic} longitude of the Sun at a meteor instant, hence to split the data sample we just used the catalog ordinal numbers of orbits: the even-numbered orbit entered sample CAMSA, the odd-numbered entered sample CAMSB. 
\begin{figure*}
\vbox{
\centerline{
\hbox{
\includegraphics[width=0.90\textwidth]{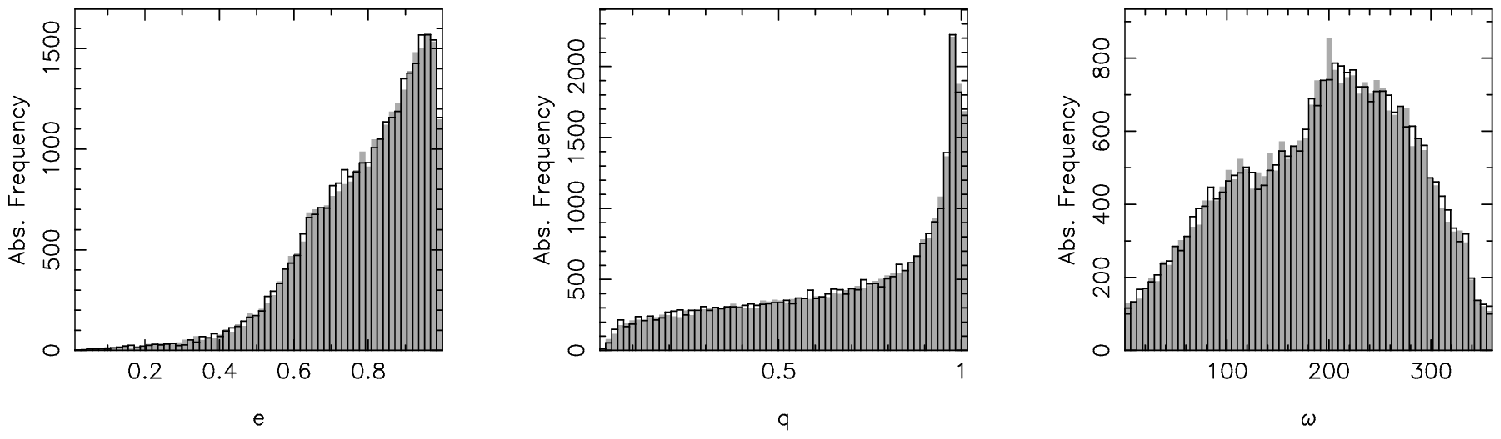}  
}
}
\centerline{
\hbox{
\includegraphics[width=0.60\textwidth]{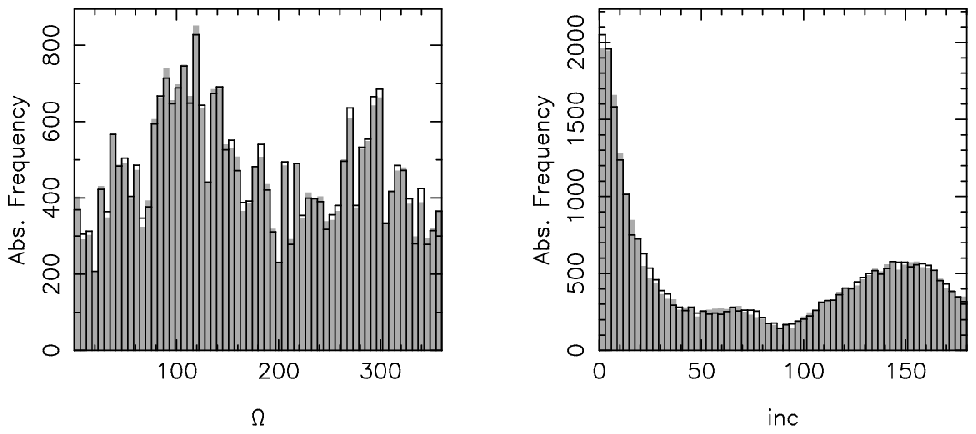} 
}
}
}
\caption{ Marginal histograms of the osculating orbital elements of the observed subsamples CAMSA and CAMSB (in gray). Each sample consists of $29014$ orbits, the number of bins is $62$.
The distributions display a clear statistical similarity. Except for one small surplus seen  in the $\omega$ histogram,  the only visible differences are the statistical fluctuations.
}
\label{fig:HiCAB}
\end{figure*}
\begin{figure*}
\vbox{
\centerline{
\hbox{
\includegraphics[width=0.80\textwidth]{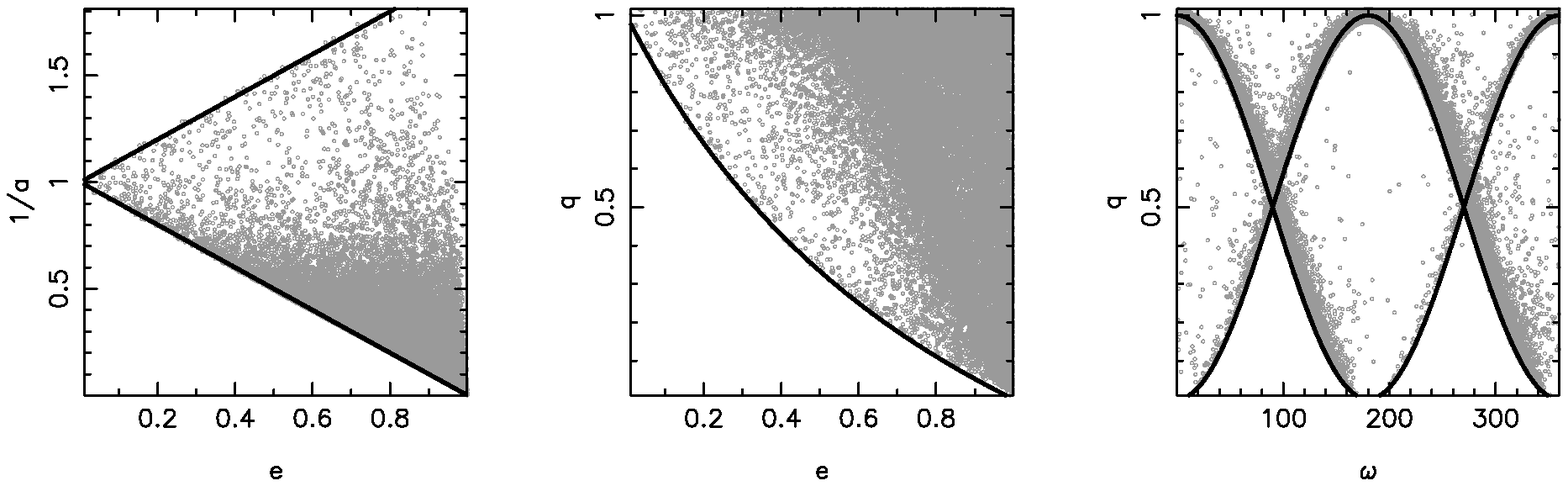}  
}
}
\centerline{
\hbox{
\includegraphics[width=0.80\textwidth]{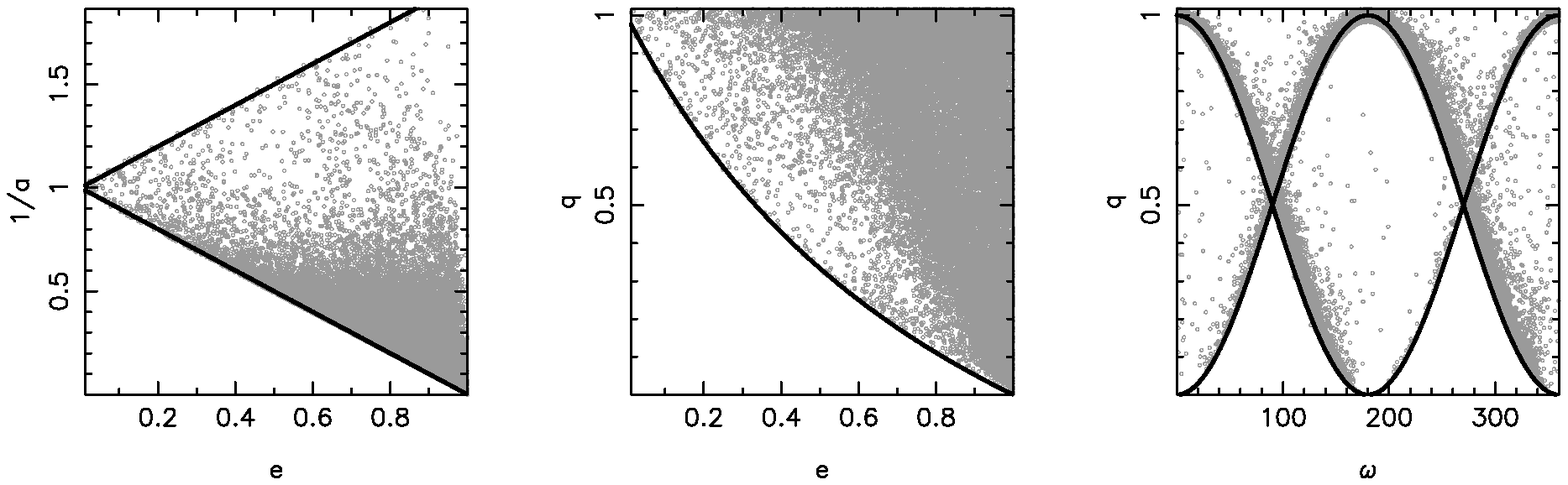} 
}
}
}
\caption{Distributions of $29014$ sporadic video meteoroids (top -- CAMSA and bottom -- CAMSB subsample) on the $q$-$e$, $(1/a)$-$e$, and $\omega$-$q$ planes. Due to the observation selection effect, not all combinations of  $[q,e]$,  $[(1/a),e]$,  or $[\omega,q]$ are observable from the surface of the Earth.
}
\label{fig:2DCAB}
\end{figure*}
\begin{table}
\scriptsize
\footnotesize
\caption[]{
Results of the quantitative comparison of $29014$ observed and synthetic meteoroid orbits.
The first section contains the results obtained for the CAMSA and CAMSB subsamples.  The second section contains the results determined for synthetic samples generated using CAMSA orbits (see text).
The $\chi^2$ distances were calculated using $\omega$-$q$ histograms with $62$x$62$ bins. Subsequent columns contain ${OS_N}$  and  ${NN_N}$ statistics.
}
\begin{center}
\begin{tabular}[t]{lccc}
\hline
\hline
\multicolumn{1}{l}{Orbital sample} & \multicolumn{1}{c}{$\chi^2$ distance} & \multicolumn{1}{c}{${OS_N}$} &  \multicolumn{1}{c}{${NN_N}$}  \\
\hline
CAMSA  29014     &   -       &      -    &     0.097  \\
CAMSB  29014     &    760    &   0.096   &     0.096  \\
\hline
Method E        &    7424   &  0.119     &    0.120  \\
KD, h=10        &    6499   &  0.094     &    0.103  \\
KD, h=0.1       &    473    &  0.048     &    0.051  \\
KD, non-scalar  &    3103   &   0.089    &    0.096  \\
\hline
\label{tab:comparison2}
\end {tabular}
\end {center}
\normalsize
\end {table}
\begin{figure*}
\begin{center}
\centerline{
\vbox{
\hbox{
\includegraphics[width=0.33\textwidth]{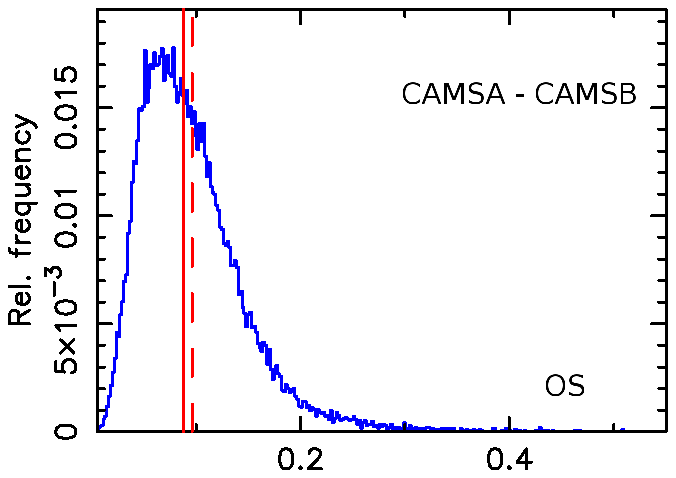}       \hspace{-0.20cm}
\includegraphics[width=0.33\textwidth]{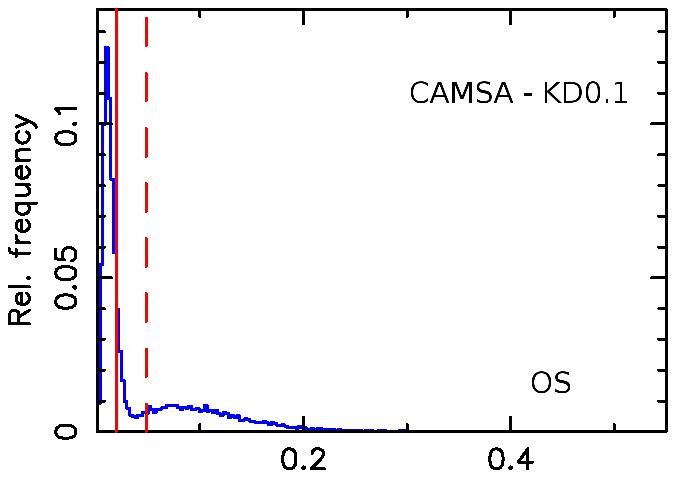}    \hspace{-0.2cm}
\includegraphics[width=0.33\textwidth]{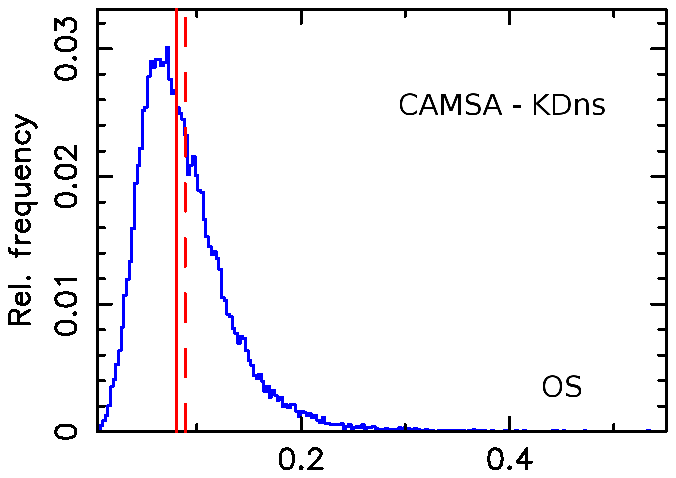} 
}
\hbox{
\includegraphics[width=0.33\textwidth]{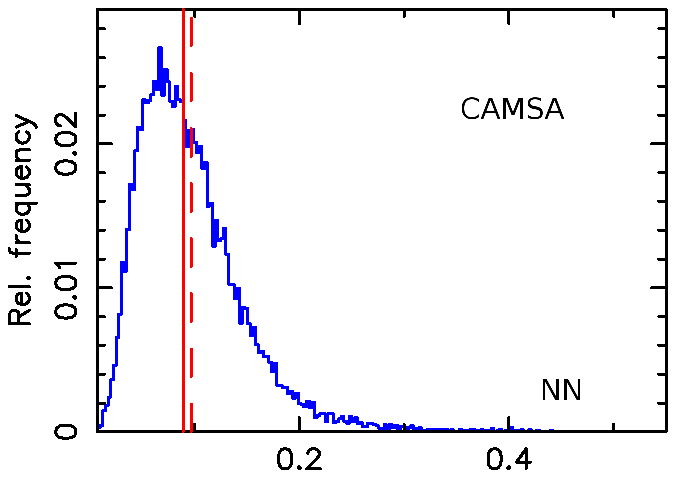}    \hspace{-0.20cm}
\includegraphics[width=0.33\textwidth]{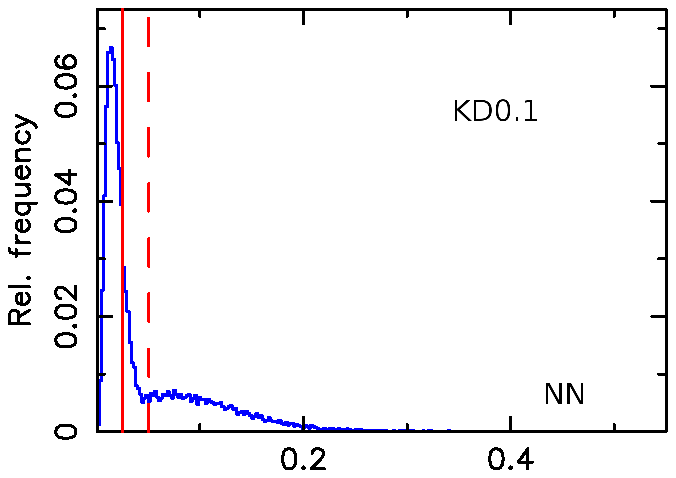}    \hspace{-0.20cm}
\includegraphics[width=0.33\textwidth]{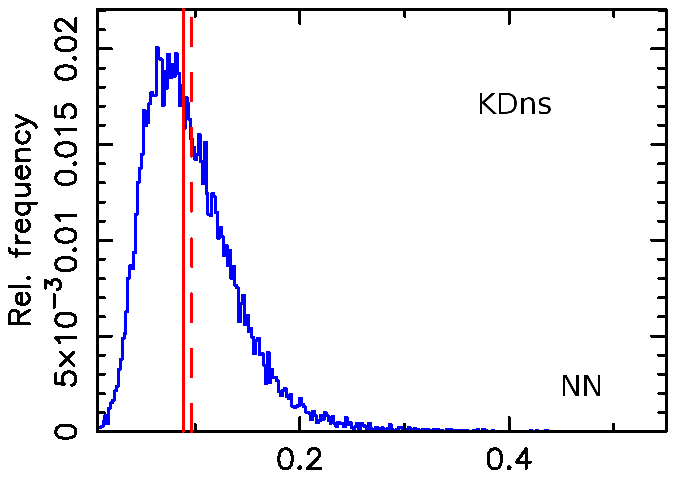} 
}
}
}
\end{center}
\caption{Histograms of the values used to calculate the  ${OS_N}$  (top) and ${NN_N}$ (bottom) statistics.
The dashed red lines correspond to the ${OS_N}$ and ${NN_N}$ mean values, the continuous lines correspond to the median values.
The $OS$ histograms were produced using the  observed CAMSA and CAMSB orbits (left), the CAMSA and KD0.1 (middle), and the CAMSA and KDns samples (right). The $NN$ histograms correspond to CAMSA, KD0.1, and KDns samples, respectively. For the KD0.1 synthetic data, the histograms are considerably different from the remaining ones.
}
\label{fig:histSONN}
\end{figure*}

The CAMSA and CAMSB subsamples were taken from the same 'population' of the observed orbits, hence their statistical properties should be very similar. 
{ In Fig.~\ref{fig:HiCAB} we see that their marginal  distributions of the orbital elements display a clear statistical similarity. This similarity is also visible in the 2D distributions in Fig. \ref{fig:2DCAB}. This figure shows the effects of observational selection that correlate some elements of the meteoroid orbit. 
Because a typical meteoroid can be observed only in the form of a meteor phenomenon in the Earth's atmosphere, an approximate relationship between its orbital elements $q$, $e$, and $\omega$ is often used \citep[see, e.g.,][]{1997A&A...320..631J, 2012MNRAS.420.2546B}, namely
\begin{equation}
q={{R_E\,(1 \pm e\cos\omega)} \over {1+e}},
\label{row:03b}
\end{equation}
where $R_E$ in astronomical units is the  heliocentric distance of a meteoroid at its moment of collision with Earth.  
The sign at the $e\cos \omega$ is positive for the negative geocentric ecliptic latitude of a meteor 
radiant, \mbox{$\beta_G<0$}.
Assuming a circular orbit of the Earth ($R_E$=$1$), the relationship $q=f(\omega)$ has been drawn in Fig. \ref{fig:2DCAB} in the form of continuous lines.
Due to the nonzero eccentricity of the Earth's orbit, $R_E$ belongs to the interval
\begin{equation}
0.9833 \leq  R_E \leq 1.0167,
\label{row:03c}
\end{equation}
and as a result, the points corresponding to the observed meteoroids surround the curve for $R_E=1$.

If we ignore the size of the Earth, that is, if we assume that meteors are observed at the center of the Earth, then Eq. (\ref{row:03b}) is exact.
Violations of Eq. (\ref{row:03b}) and Condition (\ref{row:03c}) do not happen often \citep[see][]{2010pim8.conf...91J}, but in $58029$ sporadic meteoroid orbits from the CAMS catalog they occurred $\sim$$2000$ times.

Another consequence of the observational selection concerns the values of the meteoroid's orbital elements $e,q,$ and $1/a$. As we see in Fig. \ref{fig:2DCAB},  certain combinations of the values of these elements are unacceptable for meteoroids observed from the Earth's surface. For the given values of $e$ and $R_E$,  the limit values for $q$ and $1/a$ are
\begin{eqnarray}
 q_{max}=R_E ,   \nonumber \\
 q_{min}=R_E\cdot(1-e)/(1+e), \nonumber \\
\label{row:04}
 (1/a)_{max}=1/R_E\cdot(1-e), \\
 (1/a)_{min}=1/R_E\cdot(1+e), \nonumber
\end{eqnarray}
where $R_E$ belongs to interval (\ref{row:03c}). Equations (\ref{row:04}) describe the boundary curve $q(e)$ and two limits for $(1/a)$  seen in Fig. \ref{fig:2DCAB}.

The similarity of the 1D and 2D distributions, shown in Figs. \ref{fig:HiCAB} and \ref{fig:2DCAB}, confirms our assumption that both CAMSA and CAMSB samples represent statistical properties of the same population.
}
Therefore, for these samples we calculated  the values of the statistics used, and they were applied as the reference values when the synthetic data were evaluated.
We repeated the calculations of the $\chi^2$ distance,  and ${OS_N}$ and ${NN_N}$ statistics for reduced{ synthetic samples of $29014$ orbits} (see Table~\ref{tab:comparison2}). 
To find the $\chi^2$ distances, we used $\omega$-$q$ histograms with $62$x$62$ bins, and  the number of bins was calculated according to the earlier mentioned Rice Rule.  

In Table~\ref{tab:comparison2}, the $\chi^2$ values are two to three 
times  smaller than those provided in Table~\ref{tab:comparison1}. In Table~\ref{tab:comparison2}, the largest $\chi^2$ value was calculated for the ME sample, the smallest one corresponds to the KD0.1 sample. Moreover for the ${NN_N}$ statistics, the ME sample shows the large difference compared to the observed orbits. However, while it is not the largest one, the largest difference  was produced by the KD0.1 method. \citet{2017Icar..296..197V} interpret the ${OS_N}$ and ${NN_N}$ as the indicators of statistical similarity between the synthetic and observed samples.  
However, in comparison with the reference values, the small values ${OS_N}=0.048$ and ${NN_N}=0.051$ given in Table~\ref{tab:comparison2} do not mean that the KD0.1 sample is statistically very similar to the observed one. This is obvious when we look at Fig.~\ref{fig:histSONN}. The $NN$ histograms for the observed and KD0.1 samples should look similar, but  here the discrepancy is considerable, and we have to admit that we do not know what the reason is. In addition, in Table~\ref{tab:comparison2}, for the KD0.1 method the value of the $\chi^2$ distance is too small. The $\chi^2$=$473$ is smaller than $\chi^2$=$760$ calculated for the CAMSA and CAMSB, two samples drawn from the same 'population'. 

The ${OS_N}$ and ${NN_N}$ statistics do not give an unambiguous  assessment of the quality of the synthetic orbit generator. As we can see in Table \ref{tab:comparison2}, if we use ${OS_N}$ statistics, then method KD10 turns out to be the best; if we choose ${NN_N}$ statistics, then method KDns is the best.

In conclusion, we believe that the application of the $\chi^2$ distance, and ${OS_N}$ and ${NN_N}$ statistics  only may not be  sufficient to decide which synthetic sample more adequately represents the observed data. 
\subsection{Extended comparison and discussion}
To assess whether the synthetic orbital data sets are of acceptable quality, in \citet{2017Icar..296..197V} the authors used the $\chi^2$ distance calculated for $\omega$ and $q$ only, so using only a part of the orbital sample, and as we have shown such a limited approach may not always be applicable.
We think that to evaluate the properties of a synthetic orbit generator, one should perform a more comprehensive test that seeks to establish the statistical similarity of two data sets using all five orbital elements. Moreover, from  a practical point of view it is essential to compare the results obtained with different synthetic samples used in a specific issue, for example, in a cluster analysis.

In the next section, we present the results of the extended tests of the observed and synthetic data. We also present the results of calculations of the thresholds of orbital similarity, which are key parameters of each cluster analysis using the meteoroid data.
\subsubsection{$\chi^2$ and K-S statistical tests}
Can we prove that two orbital data sets are consistent with a single distribution function? As far as we know,  there is no affirmative answer to this question. Hence, we do not have a  ready-for-use solution.
However, using statistical tests, we can compare the 1D and 2D distributions of the orbital element samples. 
Hence, in this study, we wanted to find which available statistical tests can be helpful in assessing the quality of the orbital sample generators.  We used the $\chi^2$ test (not to be confused with the $\chi^2$- distance given by Eq. (\ref{Pele})) and two K-S tests. All tests were performed using the subroutines taken from the `Numerical Recipes' package \citep{2002numrec.book.....P}, namely: {\em chstwo}, {\em kstwo}, and {\em ks2d2s}. Each subroutine returns an estimate of the significance {\em prob} parameter, which determines the level of consistency between compared distributions. The {\em prob} takes values in the range $[0,1]$, however, concerning the critical value of this parameter, in \citet{2002numrec.book.....P} the authors only note that a small value  of {\em prob} indicates a significant difference between the compared distributions. Hence, to estimate what values of {\em prob} can be expected, we performed the $\chi^2$ and K-S tests using the observed CAMSA and CAMSB samples.
\subsubsection*{One-dimensional $\chi^2$ tests} 
Making use of the 1D $\chi^2$ test, we compared the marginal distributions of $e,q,\omega,\Omega,i$ orbital elements of the CAMSA sample with CAMSB and synthetic samples. 
The results of the tests are collected in Table~\ref{tab:chi-KS1}. We tested five methods, and their algorithms are described in the Appendix.  
\begin{table}
\footnotesize
\caption[]{
Results of the 1D $\chi^2$ test (first section) and 1D K-S test (second section) for marginal distributions of the orbital elements. We compared the orbits from the observed CAMSA sample with the orbits generated by the KD10, KDns, ME0, ME1, and ME4 methods. The results of a comparison of the CAMSA and CAMSB samples serve as a reference for the other tests.
}
\begin{center}
\begin{tabular}[t]{lcccccc}
\hline\hline
            &       CAMSB&          KD10&       KDns&         ME0&         ME1&          ME4\\
            \hline
  q         &       45.7&       0.0&        0.0&        0.0&      0.0&      0.0\\ 
  e         &       35.3&       0.0&        0.0&       67.4&     94.0&     97.1\\ 
  i         &       68.1&       4.0&       35.7&      100.0&     96.4&     98.3\\ 
$\omega$    &       80.2&       0.0&        0.7&       36.9&     20.2&    100.0\\ 
$\Omega$    &      100.0&      98.4&       99.9&       99.9&    100.0&    100.0\\ 
\hline
           &     CAMSB&       KD10&      KDns&       ME0&       ME1&          ME4  \\
\hline
  q        &    33.1&         0.0&        0.0&        0.0&        0.0&           0.0\\  
  e        &    98.4&         0.0&        0.0&       18.0&       14.4&          65.2\\ 
  i        &    78.8&         0.0&        0.0&       48.2&       92.8&          50.8\\ 
  $\omega$ &    31.1&         0.2&        0.8&       57.5&       87.7&          88.2\\ 
  $\Omega$ &    95.9&        11.4&       87.2&       35.7&       26.4&          58.9\\ 
\hline
\label{tab:chi-KS1}
\end {tabular}
\end {center}
\normalsize
\end {table}

As can be seen in Table 3, the comparison of the marginal distributions of the perihelion distances $q$, gave negative results for all synthetic samples. This is not surprising in the case of ME methods, where $q$ is not generated using its marginal distribution of the observed sample. 
In each of the ME methods the perihelion distance was calculated using Eq. (\ref{row:03b}), substituting  the generated values of $e$, $\omega$ as well as the heliocentric distance $R_E$ of Earth at the moment of time of the meteor observation.
\\
For the remaining orbital elements, the results of the $\chi^2$ test are usually  more favorable for ME methods. This is due to the fact that KD orbits are not generated on the basis of the marginal distributions. Similarly, one can explain why the {\em prob} values for  $e, i,$ and partly for $\omega$ tests are lower for the CAMSB sample than for synthetic orbits generated by ME methods. The CAMSA and CAMSB samples were obtained based on the sorted values of the ecliptic longitudes of the Sun at the meteor instant (see Sect. \ref{camsab}). This also explains the highest {\em prob} value ($100$\%) obtained for the test of $\Omega$ distributions of the CAMSA and CAMSB samples.

\begin{figure*}
\begin{center}
\centerline{
\vbox{
\hbox{
\includegraphics[width=0.33\textwidth]{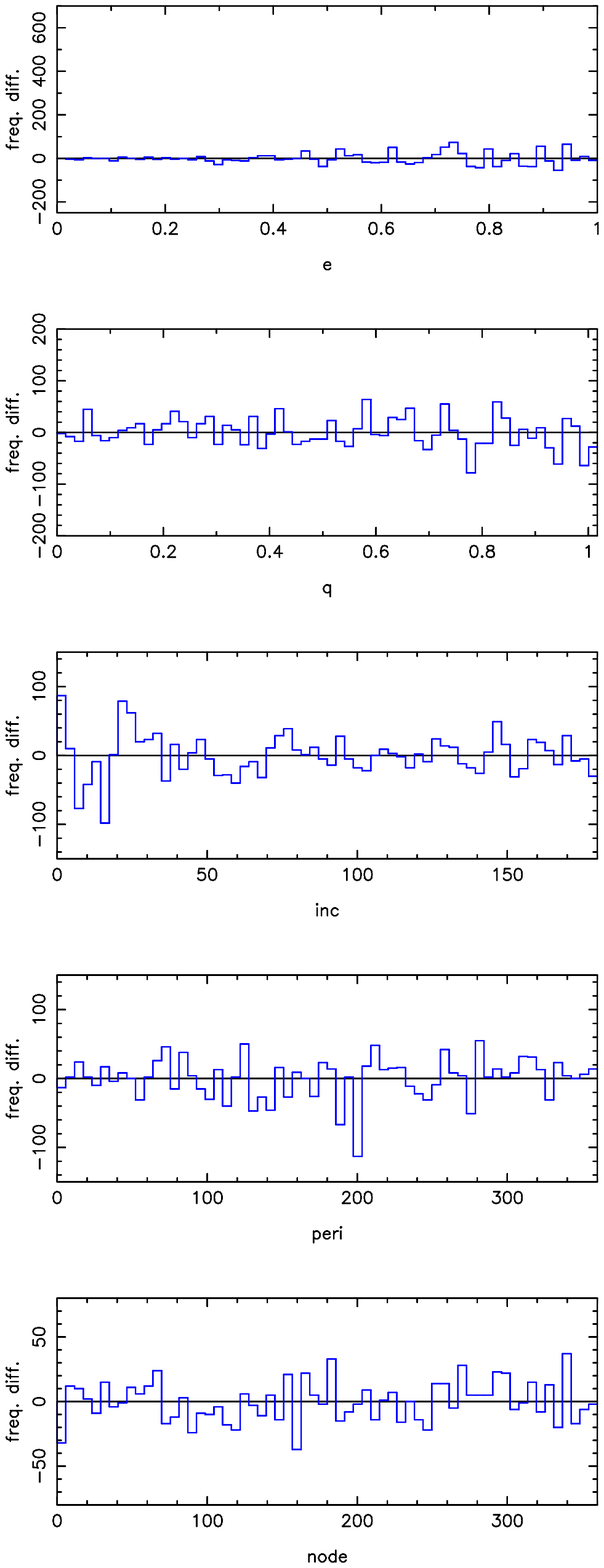}       \hspace{-0.0cm}
\includegraphics[width=0.33\textwidth]{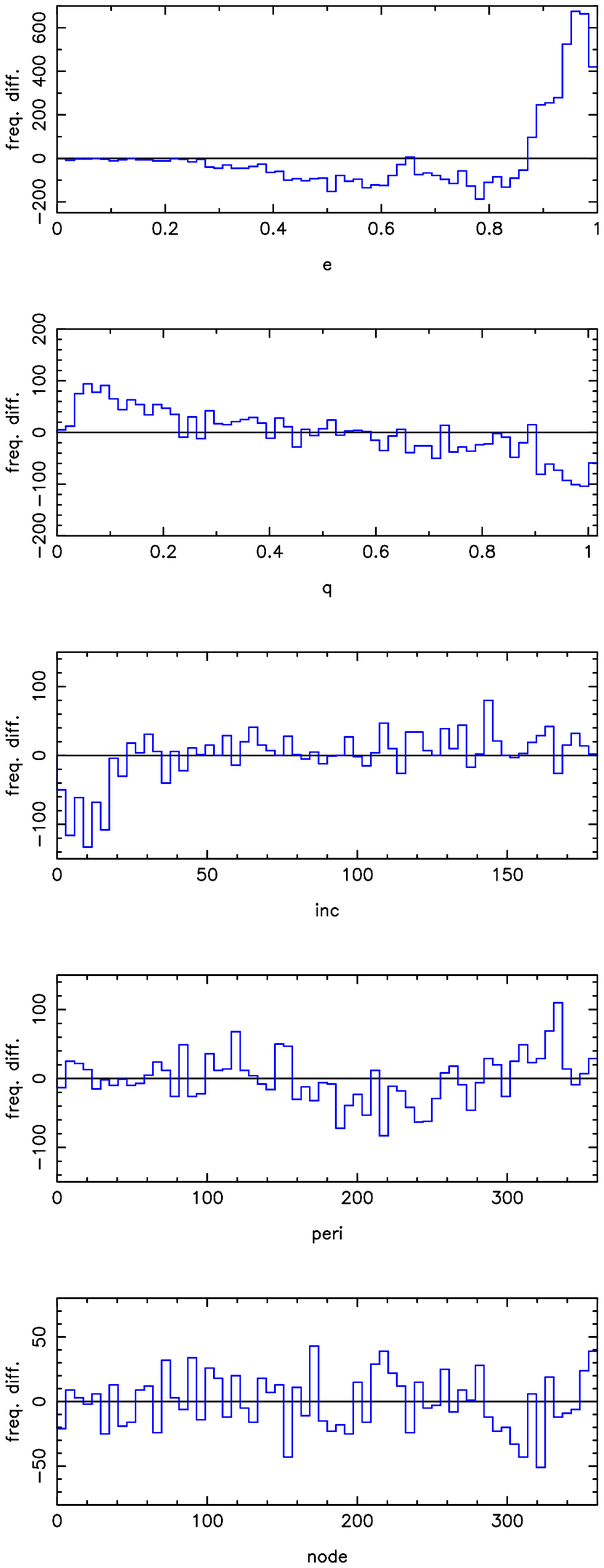}    \hspace{-0.0cm}
\includegraphics[width=0.33\textwidth]{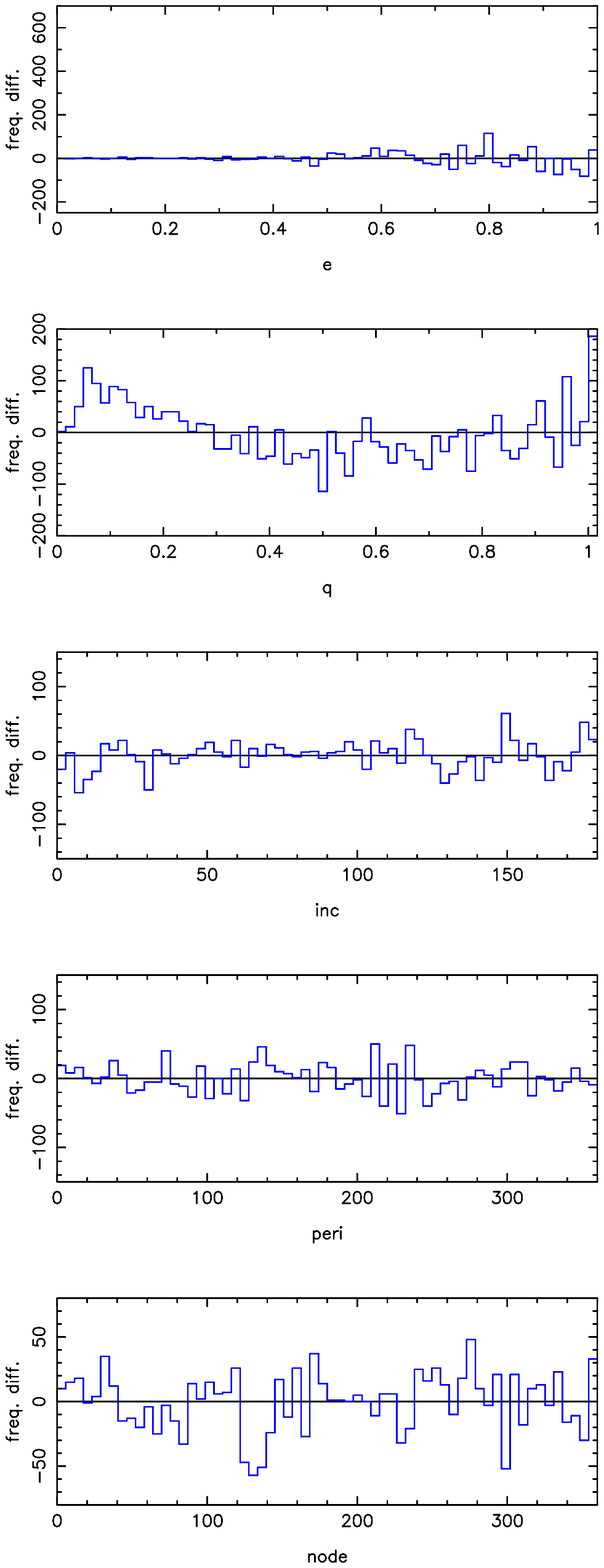}  
}
}
}
\end{center}
\caption{Plot of the differences of the absolute frequencies of the marginal distributions of orbital elements of the observed and synthetic data. From the left, in the columns we have the differences between CAMSA  and CAMSB data, between CAMSA and KDns data, and between CAMSA and ME4 data. Positive values indicate the prevalence of frequencies in a given bin for CAMSA data.  
}
\label{fig:hist-OC}
\end{figure*}
In Fig. \ref{fig:hist-OC} we plotted the differences of frequencies corresponding to the same bins of the histograms of marginal distributions of the orbital elements. The left column illustrates the differences between the CAMSA and CAMSB subsamples, the remaining columns show the differences between the CAMSA and KDns and CAMSA and ME4 samples. Except for a few cases, the differences are essentially equivalent to each other. However, the KDns method does not cope with generating orbits with large eccentricities. Similar problems arise for small and large perihelion distances, and for small inclinations. In case of the ME4 method, the same problems arise for small and large perihelion distances. The reference differences between CAMSA and CAMSB data are clearly random. This cannot be said for the differences in eccentricity, perihelion distance, and partly in inclination for the KDns sample. However, also for the ME4 method, the systematic trend is evident for the perihelion distance. Taking into account the results of the $\chi^2$-tests from Table \ref{tab:chi-KS1} and the course of differences in Fig. \ref{fig:hist-OC}, in our opinion, the ME4 method gives more consistent results than the KDns method.

\subsubsection*{One-dimensional K-S tests}

Use of binned data involves a loss of information, however. Additionally, there is often considerable arbitrariness as to how the bins should be chosen. The results given in the first section of Table \ref{tab:chi-KS1} were obtained assuming the following ranges for the observed and synthetic orbital elements: $e\in[0,1)$, $q\in[0,1.068)$, $i \in[0.180)$, $\omega \in[0,360)$ and $\Omega \in [0,360)$. For each orbital element, these intervals were divided by $62$, and the number of bins was calculated by the Rice Rule.   

To avoid the problem of the influence of the number { of bins},
we applied the K-S test to the unbinned data of $e,q,\omega,\Omega,i$ orbital elements. We used the 1D K-S test, and more particularly, we used the {\em kstwo}  subroutine taken from the `Numerical Recipes' package \citep{2002numrec.book.....P}.
The results of the K-S tests are given in the second section of Table \ref{tab:chi-KS1}. Like the earlier $\chi^2$ tests, the 1D K-S test for $q$ gave negative results for all synthetic samples. Furthermore, like the previous test the K-S test shows that the ME methods give results closer to the observed sample than the KD methods.

\subsubsection*{Two-dimensional tests}

We describe the results of a comparison of the observed and synthetic samples applying the $\chi^2$-distance (Eq. (\ref{Pele})) and a two-dimensional K-S test for all pairs of orbital elements. The 2D K-S test was performed using the {\em ks2d2s} subroutine taken from the `Numerical Recipes' package \citep{2002numrec.book.....P}. Results are provided in Table \ref{tab:chi-KS2}.
The first section of Table \ref{tab:chi-KS2} provides the $\chi^2$ distances between the observed CAMSA and the synthetic KD10, KDns, ME0, ME1, and ME4 samples. The results for the observed CAMSB data are for reference.  As we can see, on the basis of this table, it is impossible to establish which method gives the best orbits. The result depends on what pair of orbital elements we choose. Choosing the pair $q$-$\omega$, the ME4 method gives the result closest to the observed sample. Choosing $q$-$e$, KDns is the best. Choosing the pair $q$-$i$ we see that for the KDns the $\chi^2$ distance is smaller than for the reference CAMSB sample, and this indicates that KDns gives orbits that are too similar to the observed sample. Hence, for the pair $q$-$i$, the KD10 method takes precedence. The same is true for the pair $q$-$\Omega$, therefore the ME4 method takes precedence. 
In the case of KDns method, for five orbital element pairs, the $\chi^2$ distances prove to be smaller than for the CAMSB data, for the KD10 method three times, respectively. On the other hand, this was not the case with all ME methods. 

The results of the 2D K-S tests for all pairs of orbital elements are given in the second section of Table \ref{tab:chi-KS2}.
As one can see, in general the results of these tests are negative for all methods. Some exceptions are the results obtained for the pair $e$-$\Omega$  and ME0 and ME1 methods.%
\begin{table}
\footnotesize
\caption[]{
Distances of $\chi^2$  (first section) and results of 2D K-S tests (second section) for the pairs of orbital elements. The orbits from the CAMSA sample were compared with orbits generated using the KD10, KDns, ME0, ME1, and ME4 methods. The results of a comparison of the CAMSA and CAMSB samples serve as a reference for the other tests.
In the first section of the table, in each row we indicate  the values, the most (red) and least (green), differing from the result corresponding to the reference sample.}
\begin{center}
\begin{tabular}[t]{lcccccc}
\hline \hline
                   &  CAMSB &       KD10&       KDns&         ME0&         ME1&          ME4\\
\hline
  q-$\omega$       &     751&      6287&     3010&          \cz{7360}&      5512&      \zi{1743} \\
  q-e              &    1193&      2319& \zi{2148}&     \cz{4377}&      3599&      2566\\
  q-i              &    1820&      \zi{1996}&1515&     \cz{4028}&       3269&      2755\\
  q-$\Omega$       &    1915&      2229&     1120&     \cz{3560}&      2863&      \zi{1989}\\
  e-$\omega$       &    1522&      2352&     \zi{2277}&     2642&      \cz{2670}&      2656\\
  e-i              &    1402&      \zi{2115}&     2191&     2517&      2554&      \cz{2586}\\
  e-$\Omega$       &    1510&      1969&     \cz{2018}&     1585&      1527&      \zi{1521}\\
  $\omega$-$\Omega$&    1890&      \zi{1636}&     1551&     \cz{2531}&      2483&      2477\\
  $\omega$-i       &    1803&      \zi{1789}&     1732&    {3102}&      \cz{3176}&      3101\\
  i-$\Omega$       &    1992&      1623&     \cz{1617}&     \cz{2062}&      2088&      2082\\  
\hline
                  &     CAMSB&       KD10&      KDns&       ME0&       ME1&          ME4  \\
\hline
q-$\omega$        &   46.9&       0.0&        0.0&        0.0&      0.0&        0.0\\
     q-e          &   48.4&       0.0&        0.0&        0.0&      0.0&        0.0\\
     q-i          &   38.7&       0.0&        0.0&        0.0&      0.0&        0.0\\
  q-$\Omega$      &   23.8&       0.0&        0.0&        0.0&      0.0&        0.0\\
  e-$\omega$      &   31.9&       0.0&        0.0&        0.0&      0.0&        0.0\\
     e-i          &   60.0&       0.0&        0.0&        0.0&      0.0&        0.0\\
  e-$\Omega$      &   60.0&       0.0&        0.0&       \zi{26.9}&     18.1&        4.9\\
 $\omega$-$\Omega$&   34.0&       0.1&        \zi{1.6}&   0.0&      0.0&        0.0\\
  $\omega$-i      &   25.7&       0.0&        0.0&        0.0&      0.0&        0.0\\
  i-$\Omega$      &   46.9&       0.0&        0.1&        \zi{0.2}&      0.0&        0.0\\            
\hline
\label{tab:chi-KS2}
\end {tabular}
\end {center}
\normalsize
\end {table}
To sum up, it follows that assessing the quality of the synthetic orbit generator only on the basis of comparing boundary distributions can be problematic. Therefore, in the next section we describe the results of further tests based on ${NN_{N}}$ and $H_{N}$ statistics.
%
\subsubsection{Application of ${NN_{N}}$ and $H_{N}$ statistics}
\label{DHH_HN}
To compare the synthetic orbit generators, we also set the values of statistics  ${NN_N}$
and $H_{N}$. We used formulas (\ref{DNN}) and (\ref{HN}) and three  D-functions, $D_{SH},D_H$, and $\rho_1$
described in \cite{1963SCoA....7..261S}, \citet{1993Icar..106..603J}, and \citet{2016MNRAS.462.2275K} 
respectively. 
In the previous tests, we used the marginal 1D or 2D  distributions of the observed and synthetic samples.  Using ${NN_N}$ and $H_{N}$ statistics is a different approach, because to calculate them we have to use all orbital elements simultaneously. The results obtained are given in Table \ref{tab:DNN_HN}.
\begin{table}
\scriptsize
\caption[]{
Values of the ${NN_N}$ and $H_{N}$ statistics calculated for the observed CAMSA and CAMSB samples, and for the orbits generated by the KD10, KDns, ME0, ME1, and ME4 methods. The results for the CAMSA and CAMSB samples serve as a reference. The $D_{SH}$, $D_H$, and $\rho_1$ functions were used in the calculations. { In  each column, we indicate  the values, the most (red) and least (green), differing from the results corresponding to the reference samples.}
}
\begin{center}
\begin{tabular}[t]{lcccccc}
\hline\hline
 \multicolumn{1}{l}{}  & \multicolumn{3}{c}{$NN_N$} & \multicolumn{3}{c}{$H_{N}$}  \\ 
   Sample&   $D_{SH}$ & $D_H$  & $\rho_{1}$&  $D_{SH}$ & $D_H$  & $\rho_{1}$\\
\hline
CAMSA    &0.097 & 0.098 & 0.106 & 9.077 & 9.087 & 9.161 \\
CAMSB    &0.097 & 0.098 & 0.106 & 9.076 & 9.086 & 9.161         \\
KD10    &0.103 & 0.102 & {0.109} & 9.170 & 9.150 & 9.224                \\
KDns    &\zi{0.097} & \zi{0.096} & \zi{0.104} & \zi{9.098} & \zi{9.090} & \zi{9.172}              \\
ME0      &\cz{0.120} & \cz{0.124} & 0.125 & 9.308 & 9.332 & 9.356               \\
ME1      &\cz{0.120} & \cz{0.124} & \cz{0.126} & \cz{9.320} & \cz{9.338} & \cz{9.365}            \\
ME4      &0.113 & 0.114 & 0.121 & 9.245 & 9.243 & 9.324         \\
\hline
\label{tab:DNN_HN}
\end {tabular}
\end {center}
\normalsize
\end {table}

This time, the values of ${NN_N}$ and $H_{N}$ for the KD methods proved to be closer to the values for the observed samples than for the ME methods.
The KDns gave the closest values for all D-functions, the ME4 method gave the closest values among the ME methods. The values of ${NN_N}$ and $H_{N}$ depend on the D-function used in the calculations, which is understandable because these functions are not mutually equivalent. For the  $\rho_1$ function, the results were always greater than for the remaining ones. For the $D_{SH}$ and $D_H$ functions, the results are similar to each other, but in general for the former they are slightly smaller.

It might seem that both statistics are equivalent to each other and differ only by a scaling factor. In general, however, this is not the case, and this can be seen when comparing the $NN_N$ and $H_N$ values calculated for the CAMSB and KDn1 samples. For the $D_{SH}$ function, the $NN_N$ values for both these samples equal $0.097$, while for the $H_N$ they are $9.076$ and $9.098$, respectively. A similar situation occurs when we compare the results obtained using the $D_H$ function for the same CAMSB and KDns samples. The $NN_N$ statistics are $0.098$ and $0.096$ and $H_N$ are $9.086$ and $9.090$. The mean neighbor distance $NN_N$ for the CAMSB sample may be smaller { or} greater
than for a given KDns sample, while the entropy of CAMSB sample may be smaller { or} greater than for the same KDns sample. 

It should be recalled that these statistics differ in their interpretation; only the $H_n$ is a measure of the entropy of a data set, which in our opinion  gives the $H_N$ statistics an important advantage over $NN_N$.
%
\subsubsection{Comparison in practical application}
\label{D-thresholds}
\begin{figure}
\begin{center}
\centerline{
\vbox{
\hbox{
\includegraphics[width=0.450\textwidth]{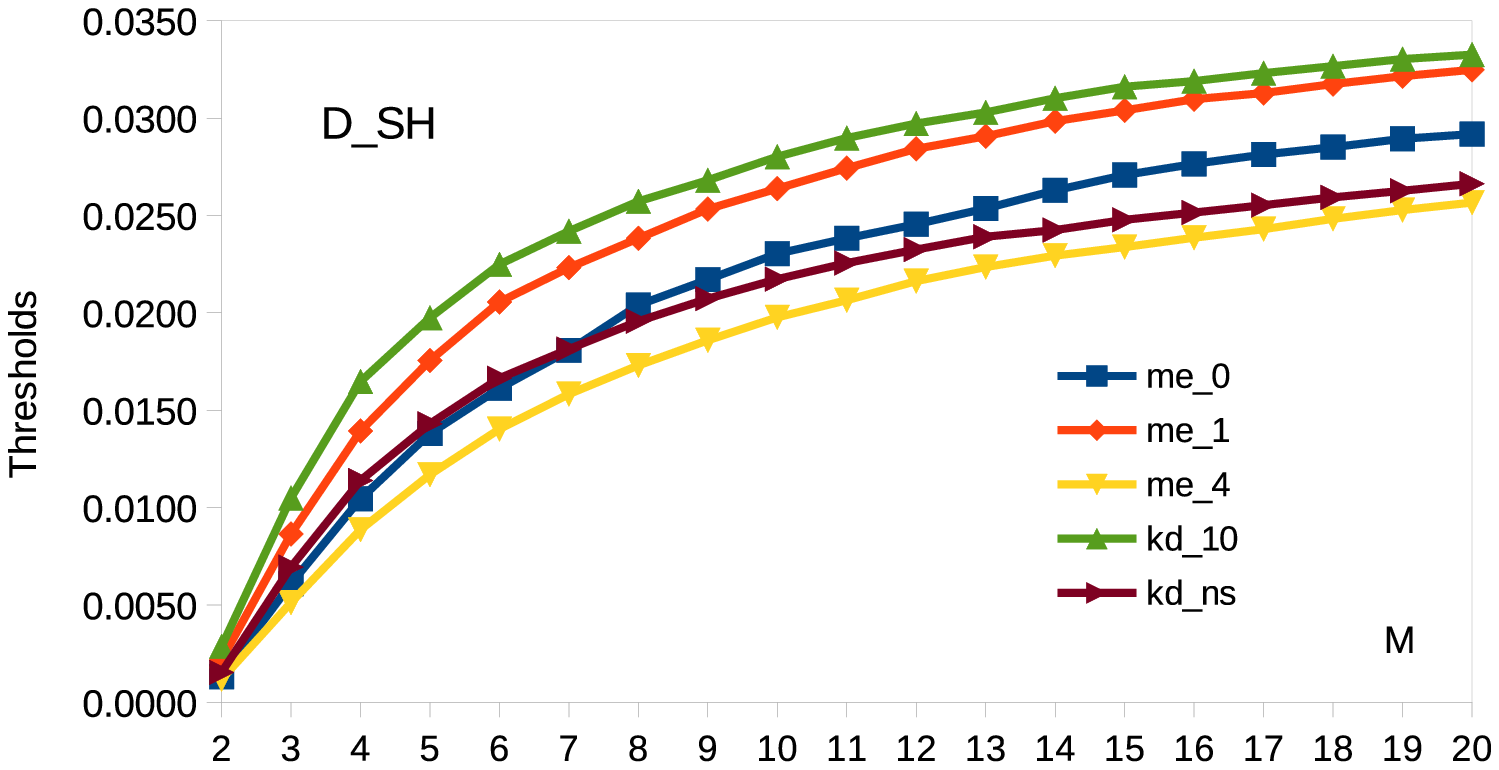}  
}
\hbox{
\includegraphics[width=0.450\textwidth]{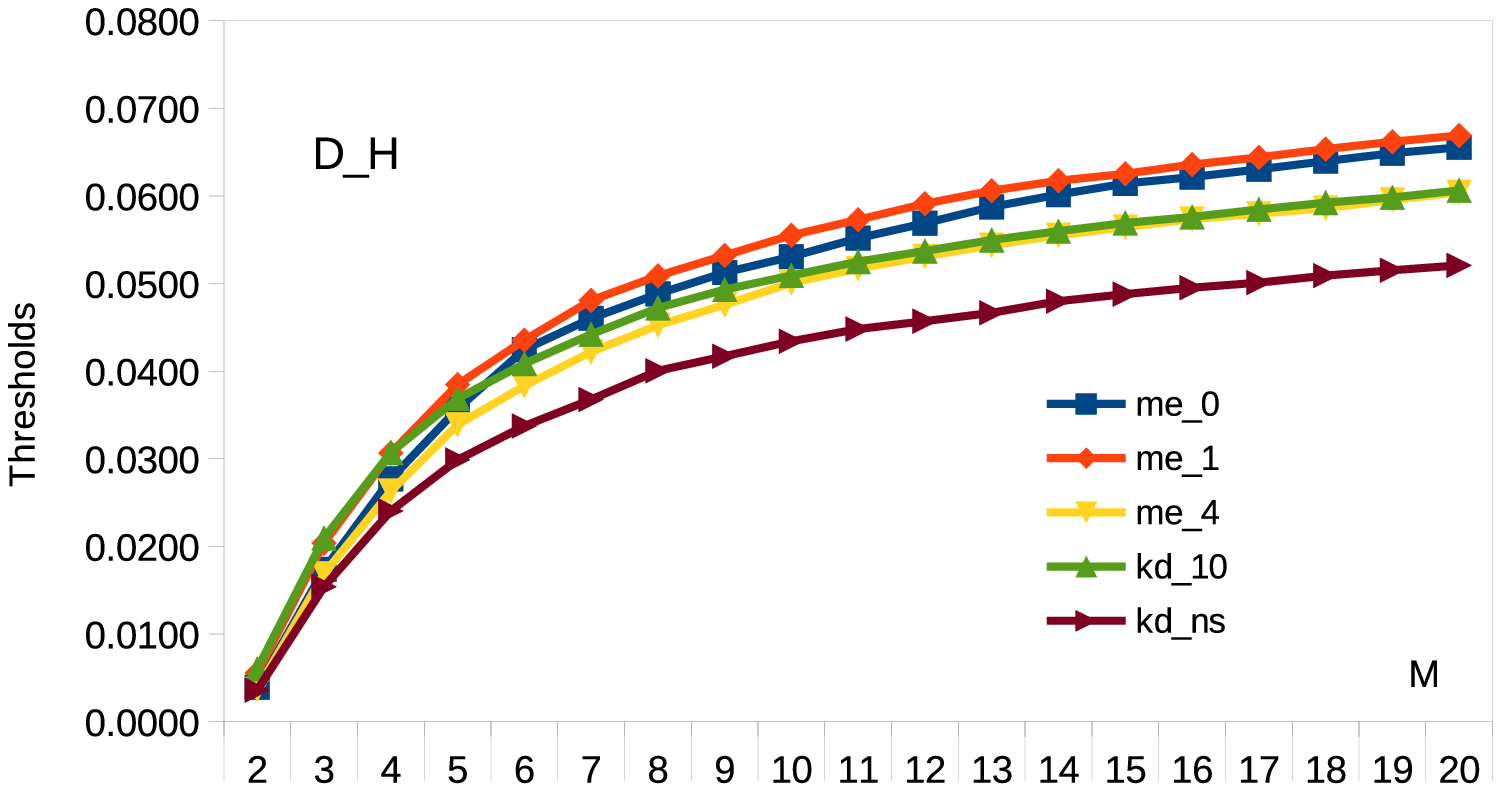} 
}
\hbox{
\includegraphics[width=0.450\textwidth]{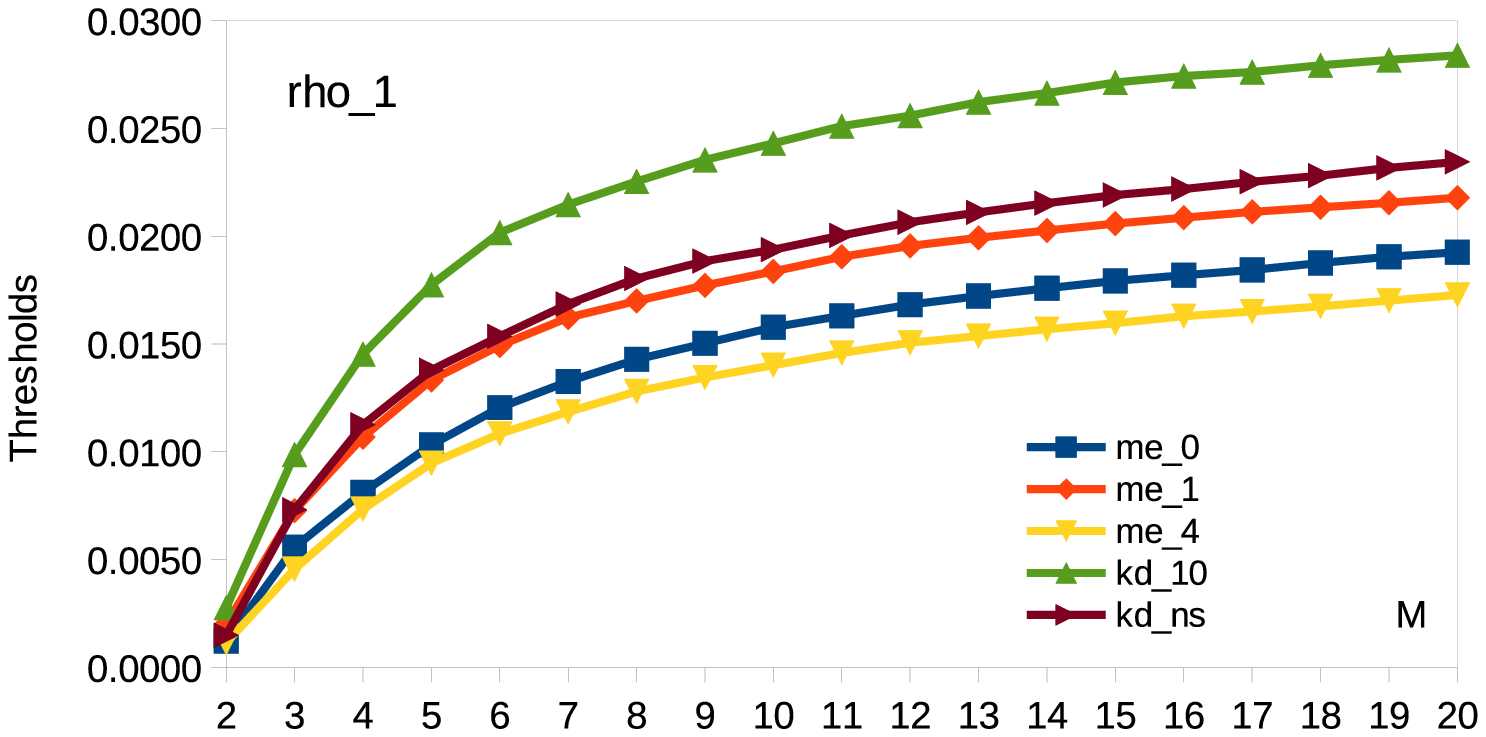}
}
}
}
\end{center}
\caption{Orbital similarity thresholds calculated for synthetic samples ME0, ME1, ME4, KD10, and KDns, using the $D_{SH}, D_H$, and $\rho_1$ functions. The thresholds were determined for groups of $M$=$2,3, ..., 20$ members, identified among $29014$ orbits.
}
\label{fig:thresholds}
\end{figure}

Both ${NN_N}$ and $H_{N}$ statistics are the point estimators, hence by using them the assessment of the quality of the generators may only be very general. We believe that a more valuable estimate of the usefulness of synthetic orbit generators can be obtained by comparing the results obtained in a specific practical application.
In this section, we present the results of calculations of the thresholds of the orbital similarity, which are key parameters of each cluster analysis. For the orbits generated by KD and ME methods, the thresholds were determined by the numerical experiment as described in \cite{2020MNRAS.494..680J}. Applying a single linkage cluster analysis algorithm, the thresholds were found for each meteoroid group of $M$=$2,3,... 20$ members, and for each of the $D_{SH}$, $D_H$, and $\rho_1$ distance functions. All obtained results corresponded to a low probability (less than 1\%) of chance clustering. The final individual thresholds $D_M$, $M$=$2,3,... 20$ were calculated as the arithmetic means of the $D_{M}$ values determined in each of the one hundred orbital samples. The final values are given in Table \ref{tab:thresholds} and their graphical illustration is shown in Fig. \ref{fig:thresholds}.%
\begin{table}
\caption{Values of the orbital similarity thresholds calculated for the synthetic samples obtained by KD and ME methods. The thresholds correspond to each group of $M$=$2,3,... 20$ members and for each distance function $D_{SH},D_{H}$, and  $\rho_1$. The provided values  are closely related to the single-linkage cluster analysis method and the size of the synthetic samples used. In this study, we used 29014 orbits.}
\begin{center}
\footnotesize
\begin{tabular}{ccccccc}
\hline\hline
\multicolumn{1}{l}{} & \multicolumn{1}{c}{M}  
                       &  \multicolumn{1}{l} {ME0} 
                       &  \multicolumn{1}{l} {ME1} 
                       &  \multicolumn{1}{l} {ME4}   
                       &  \multicolumn{1}{l} {KD10}  
                       &  \multicolumn{1}{l} {KDns}  \\ 
\hline
&2      &       0.0013  &       0.0022  &       0.0012  &       0.0029  &       0.0015  \\
&3      &       0.0061  &       0.0086  &       0.0051  &       0.0105  &       0.0069  \\
&4      &       0.0105  &       0.0139  &       0.0089  &       0.0165  &       0.0114  \\
&5      &       0.0138  &       0.0176  &       0.0117  &       0.0198  &       0.0143  \\
&6      &       0.0161  &       0.0206  &       0.0140  &       0.0225  &       0.0166  \\
&7      &       0.0181  &       0.0223  &       0.0158  &       0.0242  &       0.0181  \\
&8      &       0.0204  &       0.0238  &       0.0173  &       0.0257  &       0.0196  \\
&9      &       0.0217  &       0.0253  &       0.0186  &       0.0268  &       0.0207  \\
&10     &       0.0230  &       0.0264  &       0.0198  &       0.0280  &       0.0217  \\
$D_{SH}$&11     &       0.0238  &       0.0274  &       0.0207  &       0.0290  &       0.0226  \\
&12     &       0.0246  &       0.0284  &       0.0216  &       0.0297  &       0.0233  \\
&13     &       0.0254  &       0.0291  &       0.0224  &       0.0303  &       0.0239  \\
&14     &       0.0263  &       0.0298  &       0.0229  &       0.0310  &       0.0243  \\
&15     &       0.0271  &       0.0304  &       0.0234  &       0.0316  &       0.0248  \\
&16     &       0.0277  &       0.0310  &       0.0239  &       0.0319  &       0.0252  \\
&17     &       0.0281  &       0.0313  &       0.0243  &       0.0323  &       0.0255  \\
&18     &       0.0285  &       0.0318  &       0.0248  &       0.0327  &       0.0259  \\
&19     &       0.0289  &       0.0322  &       0.0253  &       0.0330  &       0.0263  \\
&20     &       0.0292  &       0.0325  &       0.0257  &       0.0333  &       0.0266  \\
\hline
&2      &       0.0039  &       0.0056  &       0.0037  &       0.0060  &       0.0036  \\
&3      &       0.0174  &       0.0204  &       0.0169  &       0.0209  &       0.0154  \\
&4      &       0.0277  &       0.0306  &       0.0263  &       0.0307  &       0.0240  \\
&5      &       0.0357  &       0.0385  &       0.0339  &       0.0369  &       0.0299  \\
&6      &       0.0425  &       0.0435  &       0.0383  &       0.0409  &       0.0337  \\
&7      &       0.0460  &       0.0481  &       0.0422  &       0.0442  &       0.0368  \\
&8      &       0.0488  &       0.0509  &       0.0452  &       0.0472  &       0.0401  \\
&9      &       0.0513  &       0.0532  &       0.0476  &       0.0493  &       0.0417  \\
&10     &       0.0531  &       0.0555  &       0.0501  &       0.0509  &       0.0434  \\
$D_H$&11        &       0.0552  &       0.0573  &       0.0517  &       0.0525  &       0.0448  \\
&12     &       0.0569  &       0.0591  &       0.0531  &       0.0537  &       0.0457  \\
&13     &       0.0588  &       0.0606  &       0.0544  &       0.0550  &       0.0467  \\
&14     &       0.0602  &       0.0617  &       0.0555  &       0.0560  &       0.0480  \\
&15     &       0.0614  &       0.0625  &       0.0564  &       0.0569  &       0.0488  \\
&16     &       0.0622  &       0.0636  &       0.0574  &       0.0576  &       0.0495  \\
&17     &       0.0630  &       0.0644  &       0.0580  &       0.0585  &       0.0501  \\
&18     &       0.0640  &       0.0654  &       0.0586  &       0.0592  &       0.0509  \\
&19     &       0.0649  &       0.0662  &       0.0595  &       0.0598  &       0.0515  \\
&20     &       0.0656  &       0.0669  &       0.0604  &       0.0606  &       0.0521  \\
\hline
&2      &       0.0012  &       0.0020  &       0.0012  &       0.0028  &       0.0015  \\
&3      &       0.0056  &       0.0073  &       0.0046  &       0.0099  &       0.0073  \\
&4      &       0.0081  &       0.0107  &       0.0073  &       0.0145  &       0.0113  \\
&5      &       0.0103  &       0.0133  &       0.0095  &       0.0178  &       0.0138  \\
&6      &       0.0121  &       0.0149  &       0.0108  &       0.0202  &       0.0154  \\
&7      &       0.0133  &       0.0162  &       0.0119  &       0.0215  &       0.0169  \\
&8      &       0.0143  &       0.0170  &       0.0128  &       0.0225  &       0.0180  \\
&9      &       0.0150  &       0.0177  &       0.0135  &       0.0235  &       0.0189  \\
&10     &       0.0158  &       0.0184  &       0.0140  &       0.0243  &       0.0194  \\
$\rho_{1}$&11   &       0.0163  &       0.0191  &       0.0146  &       0.0251  &       0.0200  \\
&12     &       0.0168  &       0.0196  &       0.0151  &       0.0256  &       0.0207  \\
&13     &       0.0172  &       0.0199  &       0.0154  &       0.0262  &       0.0211  \\
&14     &       0.0176  &       0.0203  &       0.0157  &       0.0266  &       0.0215  \\
&15     &       0.0179  &       0.0206  &       0.0160  &       0.0271  &       0.0219  \\
&16     &       0.0182  &       0.0209  &       0.0163  &       0.0274  &       0.0222  \\
&17     &       0.0184  &       0.0211  &       0.0165  &       0.0276  &       0.0225  \\
&18     &       0.0188  &       0.0213  &       0.0168  &       0.0279  &       0.0228  \\
&19     &       0.0191  &       0.0216  &       0.0170  &       0.0282  &       0.0232  \\
&20     &       0.0193  &       0.0218  &       0.0173  &       0.0284  &       0.0235  \\
\hline
\label{tab:thresholds}
\end{tabular}
\end{center}
\normalsize
\end{table}

For obvious reasons, we could not compare the results obtained for the synthetic orbits with the thresholds calculated for the observed orbits, hence, our comparisons relate only to synthetic orbits.
As expected, the threshold values proved to depend on the applied D-functions and the synthetic data set. Although in all cases the thresholds increased monotonically with the size $M$ of the identified group, this dependency is quite diverse. 
In Fig. \ref{fig:thresholds} we see that in the case of the $D_{SH}$ function the highest thresholds are for the KD10 orbits, the smallest one for the ME4 orbits. 
This means that if we use the $D_{SH}$ function and the KD10 orbits in the cluster analysis, then by using the obtained thresholds in the sample of observed orbits, some of the results may turn out to be artifacts, both for the number of identified streams and the number of their members. Using the ME4 orbits, we can face the opposite situation, we cannot identify certain streams, and the number of members of the identified streams may be significantly smaller. 

When the $D_H$ function is applied, the smallest thresholds are for the KDns orbits, the highest for the ME1 orbits. Finally, in the case of the $\rho_1$ function, again, the smallest values are for the ME4 orbits, the highest for the KD10 orbits. 

However, we cannot claim that any of the calculated threshold values are too large or too small. That is because we cannot compare the results obtained for the synthetic and  observed orbits here. 
We can only claim that in the case of cluster analysis, among the CAMS orbits (using a single linkage algorithm and $ D_{SH} $ or $\rho_1 $ function), the values of the orbital similarity thresholds will be the smallest if we generate synthetic orbits using the ME4 method. If we apply the $D_H$ function, the smallest thresholds will be obtained using the KDns method.
Moreover, we cannot be sure that what has been said above would also be true for the observed orbits taken from another source or if another cluster analysis were applied. 
\section{Conclusions}
We compared five methods for generating synthetic meteoroid orbits. Three of them (ME0, KD10, and KDns) had already been proposed in the literature, two additional methods (ME1 and ME4) are new. As far as possible, the synthetic orbits were compared with the orbits obtained as a result of observations, otherwise, the comparison is of a relative nature. For quantitative comparison, we applied a few tests: the $\chi^2$-distance and $NN_N$ tests used in \citet{2017Icar..296..197V}, and, implemented in this study, one-dimensional $\chi^2$ and K-S tests, as well as a two-dimensional  K-S test. In addition to the $NN_N$ statistics, to estimate a general property of the orbital sample, we used the entropy $H_N$ of the data set based on nearest neighbor distances. Finally, we did a cluster analysis among the synthetic orbits. We calculated and compared the values of the orbital similarity thresholds.  

As a result of our research, we can make a few conclusions summarized by the following points:
\begin{enumerate}
 \item Using the $\chi^2$ distance (see Eq. (\ref{Pele})) for one pair of orbital elements only, for example, $\omega$-$q$, it is not sufficient to decide which synthetic sample more adequately represents the observed data. It may be risky, as we see in Table \ref{tab:chi-KS2}. The result of the comparison clearly depends on the pair used in the test. 
 \item The same can be said about the $\chi^2$ and K-S tests performed for the marginal distributions of each orbital element. You cannot judge the quality of the orbit generator by only testing the data for a single orbit element (see Table \ref{tab:chi-KS1}).
 Nevertheless, in our study, the ME methods performed significantly better in the 1D tests than the KD methods.
 \item The 2D Kolmogorov-Smirnov test gave negative results for all compared ME and KD methods. In our opinion, it follows that the algorithms of the tested methods are not capable of generating 'realistic' meteoroid orbits. 
 \item The $NN_N$ and $H_N$ statistics gave a clear result. The values of $NN_N$ and $H_N$ for the orbits obtained by KD10 and KDns are closer to $NN_N$ and $H_N$ values calculated for the observed orbits. Both statistics are based on nearest neighbor distances, but only the $H_N$ statistics is the entropy of a data set.
 \item The results of the cluster analysis among synthetic orbits turned out to be dependent on the applied D-function of orbital similarity.  For $D_{SH}$ and $\rho_1$ functions, the smallest threshold values were obtained for the ME4 method. In the case of $D_H$ function, the smallest values were obtained for KDns orbits. By the relative comparison, we cannot decide which orbits give the thresholds as too large or which as too small.
\end{enumerate}.

The results showed us that the generation of 'realistic' meteoroid orbits is a complex issue. All methods are based on the observed data, which are heavily biased by observational selection, responsible for a number of correlations between the orbital elements. Including these correlations in an algorithm of a given method is not an easy matter.
As far as we know, these correlations are not yet well understood. Perhaps an improvement in the situation can be achieved by using the geocentric parameters of the observed meteoroids (see \citet{1999CeMDA..73...55F}, \citet{1999MNRAS.304..743V}).

The same can be said about the tests and comparisons made in our study. This is also a complex issue. They give very limited results depending on the type of the test, the orbital elements used, the D-function applied, and other factors. 

However, we were pleasantly surprised that, in the case of comparisons based on the determination of the thresholds of the orbital similarity, the results obtained, albeit different, were quite similar. Thus, at least in the case of the single linkage cluster analysis, we think that all ME and KD methods can be applied successfully.
%
\begin{acknowledgements}
The author acknowledges the anonymous referee for the comments and suggestions which improved the manuscript.
We acknowledge Peter Jenniskens for his kind decision to make the CAMS video data available to the meteor astronomers community. 
Tadeusz J. Jopek was supported by the National Science Center in Poland (project No 2016/21/B/ST9/01479).
This research has made use of NASA's Astrophysics Data System Bibliographic Services. 
\end{acknowledgements}
\bibliographystyle{aa} 
\bibliography{gene_orbit.bib} 
%
\begin{appendix} 
\section{Generation of the synthetic orbits}
%

\subsection{ME methods}
\label{ME}
\begin{figure}
\centerline{
\includegraphics[width=0.35\textwidth]{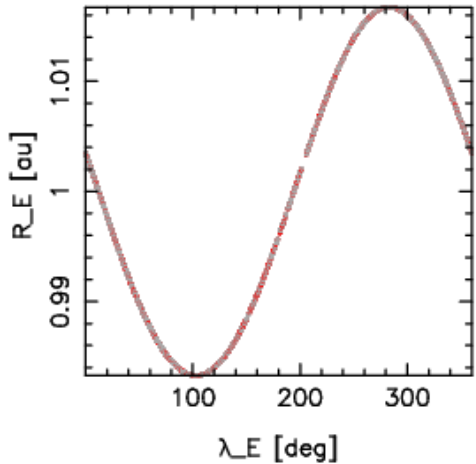}
}
\caption{Dependence of the heliocentric distance $R_E$ of Earth on the ecliptic longitude $\lambda_E$ of Earth at the time of meteor observation. Both values correspond to the meteoroids taken from the CAMSA orbital sample.
}
\label{fig:LE_RE}
\end{figure}

{ ME0} \\
The algorithm of the ME0 method  \citep[in][it was named ME]{2017P&SS..143...43J} involves the following steps: 
\begin{enumerate}
\item   For a given observed sample, obtain the histograms of the orbital elements and additionally 
        find the fraction of the orbits $\beta_{fr}$ for which the geocentric ecliptic latitudes of  the meteor radiant $\beta_G > 0 $.
\item  Generate $ e, \omega, \Omega, i$ separately by the CPD inversion method, \citep{1979generatoryZ}.
\item  Using the Earth crossing condition (\ref{row:03b}) with $R_E=1$ and the obtained $e, \omega$ calculate $q_c$;    
       choose the sign at the $e\cos \omega$ by generating the number $b$ distributed uniformly, $U(0,1)$;  
       if $b> \beta_{fr}$ , choose sign ``$-$'' otherwise choose sign ``$+$''.        
\item  Using $q_c$ and the  $e$ generated earlier, calculate the inverse of the semi-major axis $1/a_c$, if $1/a_c > (1/a)_{max}$,  
       repeat steps (2), (3), (4). ( The inverse $(1/a)_{max}$ is the maximal value in the observed orbital sample). 
\end{enumerate}
{ ME1} \\
This method is based on the ME0, however we incorporated a suggestion given in \citet{2017Icar..296..197V}, which was to set in formula  (\ref{row:03b}), the distance $R_E$  depending on the ecliptic longitude $\lambda_E$ of the Earth at the moment of time of the meteor appearance. Both $\lambda_E$ and $R_E$ values were calculated by means of the formulas taken from \citet{2012asal.book.....B}. Figure \ref{fig:LE_RE} shows that dependence $R_E$=$f(\lambda_E)$ is not random, therefore we applied its approximation
\begin{equation}
 R_E=0.016710*\sin(\lambda_E-192.211073)+1
 \label{row:RE}
.\end{equation}
Also we have modified the condition for $q_c$ and $(1/a)_c$. The calculated values had to fulfill the condition
\begin{eqnarray}
q_{min} \leq  q_c \leq q_{max},  \nonumber \\
(1/a)_{min} \leq (1/a)_c \leq (1/a)_{max},
\label{row:06}
\end{eqnarray}
where $q_{min}, q_{max}, (1/a)_{min}$, $(1/a)_{max}$ are given by formula (\ref{row:04}).\\
The algorithm of the ME1 method involves the following steps:
\begin{enumerate}
\item   For a given observed sample, obtain the histograms of $q,e,\omega,\Omega,i, \lambda_E, \beta_G$.
\item  Generate $\beta_G, \lambda_E,  e, \omega, \Omega, i$ separately by the CPD inversion method.
\item Calculate $R_E$ using formula (\ref{row:RE}).
\item  Using the Earth crossing condition (\ref{row:03b}) and obtained $R_E, e, \omega$ calculate $q_c$;    
       choose the sign at the $e\cos \omega$,  
       if $\beta_G> 0$ choose sign ``$-$'' , otherwise choose sign ``$+$'',        
\item  Using $q_c$ and the $e$ generated earlier, calculate the inverse of the semi-major axis $1/a_c$. 
\item If $1/a_c$ is outside the interval $[ (1/a)_{min}, (1/a)_{max}]$ or if $q_c$ is outside the interval $[ q_{min}, q_{max}]$, 
       repeat steps 2, 3, 4.
\end{enumerate}

\begin{figure}
\centerline{
\includegraphics[width=0.35\textwidth]{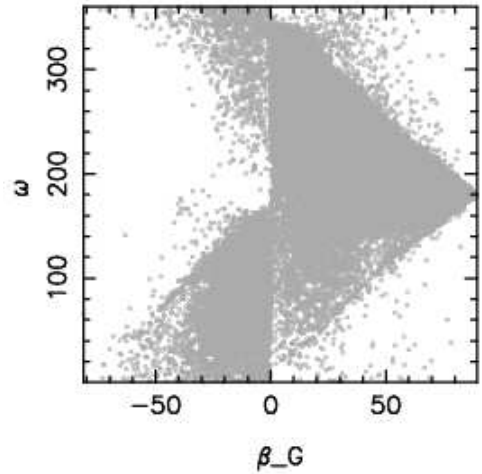}
}
\caption{Distribution of $29014$ sporadic CAMSA video meteoroids on the $\beta_G$-$\omega$ plane.  As can be seen, this plot is asymmetric due to the $\beta_G$ coordinate.}
\label{fig:beta_peri}
\end{figure}

\noindent
{ ME4} \\
The ME4 method is based on the ME1, however with one modification. We made use of the asymmetry of the $\beta_G$-$\omega$ distribution (see Fig. \ref{fig:beta_peri}). Choosing the value $\beta_G=0$, the $\omega$ values were divided into two subsamples for which $\beta_G \ge 0$ and  $\beta_G < 0$. In the method ME4, instead of the $\omega$-histogram for the whole CAMSA sample we used two histograms corresponding to $\beta_G \ge 0$ and  $\beta_G < 0$, respectively.  This solution significantly improved the test results obtained with the ME4 method. We can see this if, for example, we  compare the values for ME1 and ME4 in the first row of Table \ref{tab:chi-KS2}. 

Thus, the algorithm of the ME4 method involves the same steps as for the method ME1, except that the argument of perihelion $\omega$ was generated by the CPD inversion method using two $\omega$-histograms, depending on the sign of the generated $\beta_G$ value.
\subsection{KD methods}

The KD methods proposed in \citet{2017Icar..296..197V} for the generation of synthetic meteoroid orbits make use of the nonparametric technique. It is based on the multivariate kernel density estimation, a function which for the five dimensional problem is given as 
\begin{equation}
 \hat{f}({ O})=\frac{1}{N|{ H}|} \sum_{k=1}^{N} K_5 \left [ { H}^{-1} ({ O} - {O}_k) \right ]
 \label{kde}
,\end{equation}
where ${ O}$=$(q,e,\omega,\Omega,i)^T$, ${ O}_k$=$(q_k,e_k,\omega_k,\Omega_k,i_k)^T$, $k$=$1,2,...,N$, and $N$ is the size  the orbital sample. 
{ The letter} ${H}$ 
{ represents} the bandwidth (or smoothing) $5x5$ matrix; $K_5$ is the kernel function, symmetric multivariate density function. In \citet{2017Icar..296..197V} the authors used a Gaussian kernel.
The choice of the kernel function $K$ is not critical to the accuracy of kernel density estimators. The most important factor affecting results obtained by Eq. (\ref{kde}) is the choice of the bandwidth matrix ${ H}$.  \citet{2017Icar..296..197V} presented the results of three KD variations. In the KD10 or KD0.1 method the bandwidth $H$ was a scalar matrix with all diagonal elements equal to $10$ or $0.1$, respectively.
In the KD non-scalar method  (KDns) the bandwidth matrix is defined as
\begin{equation}
\mathbf{H} =
\left[ \begin{array}{ccccc}
 0.01 & 0.0 & 0.0 & 0.0 & 0.0  \\
 0.0  & 10  & 0.0 & 0.0 & 0.0 \\
 0.0  & 0.0 & 10  & 0.0 & 0.0 \\
 0.0  & 0.0 & 0.0 & 10  & 0.0 \\
 0.0  & 0.0 & 0.0 & 0.0 & 6 \\
\end{array} \right] ,
\end{equation}
where the values on the diagonal correspond to $q,\Omega,e,\omega,i$ respectively.
Having the multivariate estimate (\ref{kde}), it can be used to draw a synthetic sample of the orbits, for details see  \citet{2017Icar..296..197V}.
\end{appendix}
\end{document}